\documentclass{amsart}

\usepackage{natbib}

\usepackage{psfrag}
\usepackage{graphicx}
\usepackage{url}
\usepackage{amsmath,amsfonts}
\usepackage{tikz}

\newcommand\bx{{\mathbf x}}
\newcommand\bp{{\mathbf p}}
\newcommand\bu{{\mathbf u}}

\newcommand\bkappa{{\boldsymbol \kappa}}
\newcommand\bnabla{{\boldsymbol \nabla}}
\newcommand\pnabla{{\boldsymbol \nabla_\bp}}

\title{Perturbations of the coupled Jeffery-Stokes equations}
\address{Department of Mathematics, University of Missouri, Columbia MO 65211, U.S.A.}

\author{Stephen Montgomery-Smith}

\begin{document}


\begin{abstract}
This paper seeks to provide clues as to why experimental evidence for the alignment of slender fibers in semi-dilute suspensions under shear flows does not match theoretical predictions.  This paper posits that the hydrodynamic interactions between the different fibers that might be responsible for the deviation from theory, can at least partially be modeled by the coupling between Jeffery's equation and Stokes' equation.  It is proposed that if the initial data is slightly non-uniform, in that the probability distribution of the orientation has small spacial variations, then there is feedback via Stokes' equation that causes these non-uniformities to grow significantly in short amounts of time, so that the standard uncoupled Jeffery's equation becomes a poor predictor when the volume ratio of fibers to fluid is not extremely low.  This paper provides numerical evidence, involving spectral analysis of the linearization of the perturbation equation, to support this theory.

This paper differs from the published version in that it contains a corrogendum at the end.
\end{abstract}


\maketitle

\section{Introduction}

Predicting the orientation of thin fibers suspended in fluid flows with low Reynolds number finds many industrial applications, for example, when creating parts using injection molded plastics.  One method that has been widely used is to start with the assumption that Jeffery's equation \cite[][]{jeffery:23}, or some variation of it, is a good predictor of the orientation of the fibers.

However, Jeffery's equation assumes that we are dealing with a single fiber in a fluid that extends to infinity in all directions.  Thus unless the volume ratio of fibers to fluid is extremely small, it is reasonable to suppose that Jeffery's equation will require some modification.  And indeed experiments performed for large numbers of fibers in shear flows show that the prediction from Jeffery's equation is somewhat inaccurate:
\begin{enumerate}
\item The eventual steady state disagrees with the prediction from the Jeffery's equation.  Jeffery's equation only predicts a steady state if the aspect ratio of the fibers is infinite ($\lambda=1$), but a steady state seems to be observed experimentally even if the aspect ratio is finite.
\item Jeffery's equation for fibers with finite aspect ratio ($|\lambda|<1$) predicts that the fiber orientation is periodic in time (this is referred to as Jeffery's `tumbling'), but this is not seen in experiments \cite[see, for example,][]{anczurowski:67}.
\item Thirdly, the rate of any alignment is much slower than predicted.
\end{enumerate}
The first and second points were successfully countered by adding diffusion terms to Jeffery's equation \cite[][]{bird:87b}, for example, the Folgar-Tucker equation \cite[][]{folgar:84}.  The diffusion term is meant to simulate an effect similar to Brownian motion, which is assumed to arise from the fibers colliding with each other.  But experimental evidence shows that the diffusion terms fail to account for the delay in alignment, and the observed times for alignment are many times longer than the theoretically obtained values \cite[see, for example,][]{nguyen:08,sepehr:04,wang:08}.

We also bring attention to the work of \cite{stover:92}.  They experimentally measured other parameters for fibers in fluids in shear flow, such as the distribution of the so called `orbit constants' for the fibers.  This provides more evidence that when there are many fibers present, experiments deviate from the predictions obtained from Jeffery's equation.

\begin{figure}
\begin{center}
\begin{tikzpicture}\begin{scope}[scale=0.5]
\draw[thick,->] (-4,4)--(4,4);
\draw (0,2) node{Low viscosity};
\draw (0,-2) node{Low viscosity};
\draw (0,0) node{High viscosity} ellipse(3cm and 1cm);
\draw (3.4641,2) [dashed,->]arc(30:-30:4cm);
\draw (-3.4641,-2) [dashed,->]arc(30:-30:-4cm);
\draw[thick,<-] (-4,-4)--(4,-4);
\end{scope}\end{tikzpicture}
\caption{A fluid of varying viscosity reacting to a shear flow.}
\label{shear}
\end{center}
\end{figure}
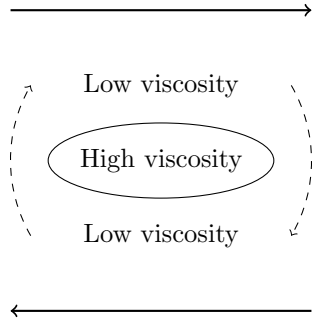

One obvious suggestion is that the deviation of experimental data from theory is caused by hydrodynamic interactions between the fibers.  That is, if a fiber is pushed by the fluid, the fiber will in turn effect the fluid.  And in turn this will cause the fluid to push other fibers differently than if the first fiber were not present.  These hydrodynamic interactions between the fibers are likely to be extremely complex.  For example, we might see some groups of fibers get close to each other and form groups that tend to move as one rather than separately.  This could, for example, be simulated by assuming that the viscosity of the suspension is higher within these groups than it is on the outside of these groups.  This alignment inside these groups is likely to be far slower.  Such a situation is illustrated in Figure~\ref{shear}.  Other examples of these complex interactions in colloids or suspensions are described, for example, in \cite{villermaux:09} and \cite{wagner:09}.  Furthermore, \cite{wagner:09} state that this kind of chaotic behavior takes place at fluid flows at much slower rates than required for the inertia of the underlying liquid to play an important factor, that is, the Reynolds number is much lower than is required for traditional turbulence to take place.

At first sight it would seem that one would have to set up large numerical experiments, which model large numbers of fibers in a Newtonian fluid, to observe this kind of effect.  However, in this paper we propose that at a good approximation of this effect might be observed without losing the assumption that the suspension is modeled well by a continuum, that is, the fluid is modeled by a simple Stokes equation.  The fibers are assumed to be smaller than the observed variations in the behavior of the suspension, so that the orientation of the fibers can be modeled by a probability distribution function.

Thus the form of Jeffery's equation we use solves for $\psi$, the probability distribution of the orientation of fibers, at each point in time and space.  This is a function $\psi(\bx,\bp,t)$ of the three variables space $\bx \in \mathbb R^3$, time $t\ge0$, and orientation $\bp\in S$, where $S = \{\bp = (p_1,p_2,p_3):|\bp|^2 = p_1^2+p_2^2+p_3^2 = 1\}$ is the two dimensional sphere.  The equations involve the velocity field $\bu = (u_1,u_2,u_3)$, which is a function of space $\bx$ and time $t$.  Associated with the velocity field $\bu$ are the Jacobian matrix $\bnabla\bu = \left({\partial u_i}/{\partial x_j}\right)_{1\le i,j\le 3}$, the deformation matrix or rate of strain tensor $\mathsf \Gamma = \bnabla\bu + (\bnabla\bu)^T$, and the vorticity matrix $\mathsf \Omega = \bnabla\bu - (\bnabla\bu)^T$.  Jeffery's equation is
\begin{gather}
\label{psi}
\frac{\partial \psi}{\partial t} + \bu \cdot\bnabla\psi = - \tfrac12\pnabla\cdot((\mathsf \Omega\cdot \bp + \lambda(\mathsf \Gamma\cdot\bp - \mathsf \Gamma:\bp\bp\bp))\psi) \\
\label{boundary-psi}
\psi = \tfrac1{4\pi} \text{ at } t=0
\end{gather}
Here $\pnabla$ denotes the gradient on the sphere $S$.

Our assertion is that if the initial state of $\psi$ is not completely isotropic, then small perturbations grow, and after a short amount of time dramatically effect the solution.  We do not produce a formula that predicts how the fibers orient, but we do cast doubt upon Jeffery's equation being able to produce a good prediction of large numbers of fibers in a Newtonian fluid.

\begin{figure}
\begin{center}

\begin{tikzpicture}\begin{scope}[scale=0.3]

\foreach \y in {-2,2} {
  \begin{scope}[yshift=\y cm]
    \foreach \x in {-8,4} {
      \begin{scope}[xshift=\x cm]
        \draw [->] (0,1.5) -- (0,1.1);
        \draw [->] (0,-1.5) -- (0,-1.1);
        \draw [->] (-1.1,0) -- (-1.5,0);
        \draw [->] (1.1,0) -- (1.5,0);
      \end{scope}
    }
    \foreach \x in {-4,8} {
      \begin{scope}[xshift=\x cm]
        \draw [->] (-0.7,1.3) -- (0.7,1.3);
        \draw [->] (0.7,-1.3) -- (-0.7,-1.3);
      \end{scope}
    }
  \end{scope}
}

\begin{scope}[yshift=2 cm]
  \draw (0,0) [xshift=-8cm,rotate=0] ellipse (0.7 and 0.1);
  \draw (0,0) [xshift=-4cm,rotate=45] ellipse (0.7 and 0.1);
  \draw (0,0) [xshift=4cm,rotate=45] ellipse (0.7 and 0.1);
  \draw (0,0) [xshift=8cm,rotate=0] ellipse (0.7 and 0.1);
\end{scope}

\begin{scope}[yshift=-2 cm]
  \draw (0,0) [xshift=-8cm,rotate=90] ellipse (0.7 and 0.1);
  \draw (0,0) [xshift=-4cm,rotate=-45] ellipse (0.7 and 0.1);
  \draw (0,0) [xshift=4cm,rotate=-45] ellipse (0.7 and 0.1);
  \draw (0,0) [xshift=8cm,rotate=90] ellipse (0.7 and 0.1);
\end{scope}

\draw [dashed] (0,6) -- (0,-4) ;

\draw (-6,6) node {High resistance} ;
\draw (6,6) node {Low resistance} ;

\end{scope}\end{tikzpicture}
\caption{How slender fibers act as `stiffeners' in elongation and shear flows.}
\label{stiffener}
\end{center}
\end{figure}
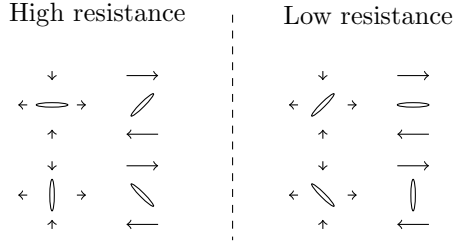

The orientation of fibers at each point effects the rheology of the suspension.  A slender fiber that is oriented parallel or perpendicular to the principle axes of the rate of strain tensor is going to act like a `stiffener' to the fluid, whereas a fiber that is oriented along a null direction of the rate of strain tensor (that is, if $\mathsf \Gamma\cdot\bp = 0$) does not hinder the flow in any way.  This is illustrated in Figure~\ref{stiffener}.  It is stated in \cite{batchelor:71,shaqfeh:90a} that if the underlying fluid is Newtonian, then the stress-strain relation for slender fibers is
\begin{equation}
\mathsf \sigma = \nu(\beta (\mathbb A:\mathsf \Gamma - \tfrac13 \mathsf I (\mathsf A:\mathsf \Gamma)) + \mathsf \Gamma) - p \mathsf I \\
\end{equation}
Here $\mathsf \sigma$ is the stress tensor, $\mathsf A$ and $\mathbb A$ are respectively the the 2nd and 4th moment tensors
\begin{gather}
\mathsf A = \int_S \bp\bp \, \psi \, d\bp \\
\label{A}
\mathbb A = \int_S \bp\bp\bp\bp \, \psi \, d\bp
\end{gather}
$\nu$ is the Newtonian viscosity that the underlying fluid would have if the fibers were absent (without loss of generality we set $\nu = 1$),  $p$ is the pressure, and $\beta$ is a dimensionless quantity that is related to the volume fraction of the fibers in the fluid.  The quantity $\beta$ represents the extent to which fibers act as `stiffeners' to the fluid motion.  We refer to the case $\beta = 0$ as the uncoupled Jeffery's equation.  The paper \cite{sepehr:04} suggests that the order of magnitude of $\beta$ could easily be as large as 50 or 100.

We assume that Reynolds number is close to zero, and so we neglect inertial terms in computing the flow of the suspension.  We also assume the suspension is incompressible.  Then the velocity field obeys the following Stokes' equation:
\begin{gather}
\label{stokes}
\bnabla \cdot \mathsf \sigma = 0 \\
\label{incompressible}
\bnabla \cdot \bu = 0
\end{gather}
Since the fluid is incompressible, the pressure $p$ is obtained implicitly, and hence without changing any of the results, we can replace $\frac13 \beta \mathsf A:\Gamma + p$ by a single scalar $q$, so that the stress-strain equation becomes
\begin{equation}
\label{stress}
\mathsf \sigma = \beta \mathbb A:\mathsf \Gamma + \mathsf \Gamma - q \mathsf I \\
\end{equation}
Next, to simplify the mathematics, we assume that the fluid occupies the whole of three dimensional space, and that there is an `ambient' velocity gradient $\mathsf U$, a three by three matrix with trace zero, so that
\begin{equation}
\label{stokes-boundary}
\bnabla\bu \to \mathsf U \text{ as } \bx \to \infty
\end{equation}
Thus we are really attempting to model the case when the fibers are much smaller than the characteristic length and width of the flow.  And we are also assuming that the perturbations whose growth in time we are calculating should have a short characteristic wavelength.  In fact, we show that the wavelength of the perturbations has no effect on the growth rate, only the direction in which oscillations take place is important to the growth rate.

\section{A heuristic argument for why alignment might be slowed down}

\begin{figure}
\begin{center}
\begin{tikzpicture}\begin{scope}[scale=0.3]
\draw[thick,->] (0,6)--(0,4);
\foreach \x in {-4,-2,0,2,4} {
  \draw (0,0) [xshift=\x cm,yshift=3cm,rotate=-60] ellipse (0.7 and 0.1);
  \draw (0,0) [xshift=\x cm,yshift=1cm,rotate=60] ellipse (0.7 and 0.1);
  \draw (0,0) [xshift=\x cm,yshift=-1cm,rotate=-60] ellipse (0.7 and 0.1);
  \draw (0,0) [xshift=\x cm,yshift=-3cm,rotate=60] ellipse (0.7 and 0.1);
}
\foreach \x in {-6,6} {
  \draw [xshift=\x cm,->] (-0.5,2) -- (0.5,2);
  \draw [xshift=\x cm,<-] (-0.5,0) -- (0.5,0);
  \draw [xshift=\x cm,->] (-0.5,-2) -- (0.5,-2);
}
\draw[thick,->] (0,-6)--(0,-4);

\begin{scope}[xshift=5cm,yshift=-6cm]
\draw[->] (-0.1,0)--(1.5,0) node[right]{$z$};
\draw[->] (0,-0.1)--(0,1) node[above]{$y$};
\end{scope}
\end{scope}\end{tikzpicture}
\caption{`Buckling' caused by elongation flow applied to a non-constant fiber orientation distribution.}
\label{buckling}
\end{center}
\end{figure}
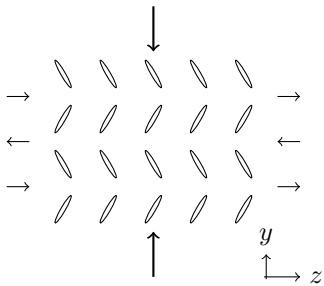

We propose that large changes in fiber orientation are created by initially small perturbations in the fiber orientation, which in a short amount of time are greatly enlarged.  An example of a growth of perturbations is a kind of `buckling' effect, illustrated in Figure~\ref{buckling}.  This shows an elongation flow where the fluid is squeezed along the $y$ axis, is expanded along the $x$ axis, and no elongation takes place along the $z$ axis, that is, $\bnabla\bu=\left[\begin{smallmatrix}G&0&0\\0&-G&0\\0&0&0\end{smallmatrix}\right]$ for some $G>0$.  The small ellipsoids represent the orientation of the perturbation, and show a small perturbation of $\psi$ of the form $\epsilon\hat\psi \sin(2\pi y/L)$.  It is reasonable to expect that this perturbation causes the fibers to `buckle,' that is, create perturbations to $\bu$ of the form $\epsilon\left[\begin{smallmatrix}0\\0\\\hat u_3\end{smallmatrix}\right] \cos(2\pi y/L)$.

This could be seen as an effect similar to placing a large number of fibers end to end, and then pushing in from both ends.  One would expect the fibers to `buckle,' but not in the $x$-direction, where they are being `guided' by the elongation flow.  Obviously this intuition in of itself is not terribly convincing, but we will show later that the mathematics does predict this effect, and indeed that this `buckling' can take place in any direction in the $yz$-plane.

This `buckling' then feeds back into the perturbation of $\psi$ causing it to grow exponentially.    We should expect similar behavior from a shear flow, because shear flow is an elongation flow at $45^\circ$ to the $x$ and $y$ axes, combined with a rotation.

However `buckling' in of itself is insufficient to cause small perturbations to become very large.  That is, over a short amount of time, buckling might cause perturbations to grow maybe two or three times.  But to really expect the perturbations to play a large role, they really need to grow ten fold or even hundred fold.

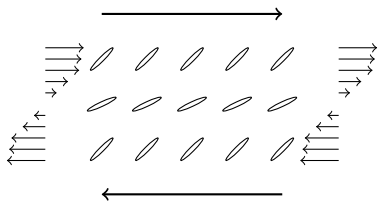
\begin{figure}
\begin{center}
\begin{tikzpicture}\begin{scope}[scale=0.3]
\draw[thick,->] (-4,4)--(4,4);
\foreach \x in {-4,-2,0,2,4} {
  \draw (0,0) [xshift=\x cm,yshift=2cm,rotate=45] ellipse (0.7 and 0.1);
  \draw (0,0) [xshift=\x cm,yshift=0cm,rotate=25] ellipse (0.7 and 0.1);
  \draw (0,0) [xshift=\x cm,yshift=-2cm,rotate=45] ellipse (0.7 and 0.1);
}
\foreach \x in {-6.5,6.5} {
  \draw [xshift=\x cm,->] (0,2.5) -- (1.7,2.5);
  \draw [xshift=\x cm,->] (0,2) -- (1.6,2);
  \draw [xshift=\x cm,->] (0,1.5) -- (1.5,1.5);
  \draw [xshift=\x cm,->] (0,1) -- (1,1);
  \draw [xshift=\x cm,->] (0,0.5) -- (0.5,0.5);
  \draw [xshift=\x cm,->] (0,-0.5) -- (-0.5,-0.5);
  \draw [xshift=\x cm,->] (0,-1) -- (-1,-1);
  \draw [xshift=\x cm,->] (0,-1.5) -- (-1.5,-1.5);
  \draw [xshift=\x cm,->] (0,-2) -- (-1.6,-2);
  \draw [xshift=\x cm,->] (0,-2.5) -- (-1.7,-2.5);
}
\draw[thick,->] (4,-4)--(-4,-4);

\end{scope}\end{tikzpicture}
\caption{Shear-banding caused by shear flow applied to a non-constant fiber orientation distribution.}
\label{shear-banding}
\end{center}
\end{figure}

To see why this much larger growth in perturbations is plausible, let us consider that we have applied a shear flow for a while, so that the fiber orientation is no longer isotropic.  When the fibers are about $45^\circ$ angle to the $x$ and $y$ axis, those fibers that are a little ahead in the alignment will have a little less stiffening effect on the fluid, and hence the shear strain will be larger on these fibers.  This will cause a positive feedback where those fibers that are a little ahead of the others will become even further ahead.  This self-reinforcing shear-banding is illustrated in Figure~\ref{shear-banding}.

We suspect that this simple model of shear banding, although plausible, is perhaps not enough to explain the very large growth of perturbations.  Rather something more complex will take place, and it seems to require all three dimensions, not just the $x$ and $y$ directions, to produce chaotic behavior.

\section{Solutions to Jeffery's equation}

The coupled system of equations we propose is a partial differential equation in six variables (three space variables, two fiber orientation variables, and the time variable).  Our task will be much easier if we can reduce this to a finite dimensional ordinary differential equation.

By \cite{lipscomb:88,dinh:84,szeri:96,montgomery-smith:09b}, it is known that if $\psi$ was ever isotropic at some time in the past, then solution to equation~\eqref{psi} is
\begin{equation}
\label{psi-B}
\psi(\bp) = \psi_{\mathsf B}(\bp) = \frac1{4\pi (\mathsf B:\bp\bp)^{3/2}}
\end{equation}
where
\begin{equation}
\label{BC}
\mathsf B = \mathsf C^T\cdot \mathsf C
\end{equation}
the matrix $\mathsf C$ satisfies the equation
\begin{equation}
\label{C-full}
\frac{\partial \mathsf C}{\partial t} + \bu\cdot\bnabla \mathsf C = - \tfrac12\mathsf C \cdot(\mathsf \Omega+\lambda \mathsf \Gamma)
\end{equation}
and equation~\eqref{boundary-psi} becomes
\begin{equation}
\label{C0}
\mathsf C = \mathsf I \text{ at } t=0
\end{equation}
Equivalently, $\mathsf B$ is a symmetric positive definite matrix with determinant one satisfying
\begin{equation}
\label{B-full}
\frac{\partial \mathsf B}{\partial t} + \bu\cdot\bnabla \mathsf B = - \tfrac12 \mathsf B \cdot(\mathsf \Omega+\lambda \mathsf \Gamma) - \tfrac12 (- \mathsf \Omega+\lambda \mathsf \Gamma) \cdot \mathsf B
\end{equation}
and equation~\eqref{boundary-psi} becomes
\begin{equation}
\label{B0}
\mathsf B = \mathsf I \text{ at } t=0
\end{equation}
Thus from now on, we assume that fiber orientation always has the form $\psi = \psi_{\mathsf B}$, given by equation~\eqref{psi-B}, for some positive definite matrix $\mathsf B$ whose determinant is one.  And perturbations of $\psi_{\mathsf B}$ are equivalent to perturbations of $\mathsf B$.

Furthermore, the 4th order moment tensor $\mathbb A$ can be calculated directly from $\mathsf B$ using elliptic integrals \cite[]{montgomery-smith:09b} \cite[see also][]{verleye:93,verweyst:98}
\begin{equation}
\label{A from B}
\mathbb A = \mathbb A(\mathsf B) = \tfrac34 \int_0^\infty \frac{s\,\mathcal S((\mathsf B+s \mathsf I)^{-1}\otimes(\mathsf B+s \mathsf I)^{-1}) \, ds}{\sqrt{\text{det}(\mathsf B+s \mathsf I)}}
\end{equation}
where $\mathcal S$ is the symmetrization of a tensor, that is, if $\mathbb B$ is a rank $n$ tensor, then $\mathcal S(\mathbb B)_{i_1\dots i_n}$ is the average of $\mathbb B_{j_1\dots j_n}$ over all permutations $(j_1,\dots,j_n)$ of $(i_1,\dots,i_n)$.

\section{Solution of the unperturbed Jeffery's equation}

The unperturbed problem is to assume that the initial data is isotropic (equation~\eqref{B0}), or at least that $\psi_{\mathsf B}$, and hence $\mathsf B$, at time $t=0$ does not depend upon $\bx$.  If $\mathsf B$ does not depend upon $\bx$, then from equation~\eqref{A} or~\eqref{A from B}, it follows that $\mathbb A$ does not depend upon $\bx$, and it becomes apparent that the solution to equations~\eqref{stokes}, \eqref{incompressible}, \eqref{stress} and~\eqref{stokes-boundary} is given by $\bnabla\bu = \mathsf U$, which does not depend upon $\bx$.  Hence from equation~\eqref{psi} or~\eqref{B-full}, it follows that $\partial\psi_{\mathsf B}/\partial t$, or equivalently $\mathsf B$, does not depend upon $\bx$.  Thus the solution to the unperturbed problem does not depend upon $\bx$ for all $t>0$.

Thus the terms $\bu\cdot\bnabla\psi$, $\bu\cdot\bnabla \mathsf B$ and $\bu\cdot\bnabla \mathsf C$ are zero, and
\begin{gather}
\label{Gamma}
\mathsf \Gamma = \mathsf U+\mathsf U^T \\
\label{Omega}
\mathsf \Omega = \mathsf U-\mathsf U^T
\end{gather}
In particular, if $\mathsf U$ is independent of $t$, then $\mathsf C$ can be computed easily by
\begin{equation}
\label{C-exp}
\mathsf C = \exp\left(-\tfrac12 t (\mathsf \Omega + \lambda\mathsf \Gamma)\right)
\end{equation}

\section{The Mathematical nature of the perturbations}

The perturbation we propose is to apply the replacement
\begin{equation}
\psi \to \psi + \epsilon \tilde\psi
\end{equation}
and assume that both $\psi$ and its perturbation satisfy equations~\eqref{psi}, \eqref{A}, \eqref{stokes}, \eqref{incompressible}, \eqref{stress} and~\eqref{stokes-boundary}.  The perturbation $\tilde\psi$ depends upon $\bx$, even though the unperturbed $\psi$ does not.  The idea is that if $\epsilon$ is small, then one assumes that terms of order $\epsilon^2$ and higher can be ignored, and from the terms of order $\epsilon$ we form a differential equation in $\tilde\psi$.  Furthermore, this differential equation is linear in $\tilde\psi$, even though the coefficients of the differential equation may depend upon $\psi$ in a possibly highly non-linear manner.  The thinking is that if the linearized perturbations $\tilde\psi$ grow by a large amount, that one might expect the terms of order $\epsilon^2$ and higher to have a large effect on the solution.  At this point the linearization is no longer valid, and it is reasonable to assume that after this point the solution becomes chaotic, or at least differs significantly from the unperturbed solution.

We can write the perturbation as a Fourier transform
\begin{equation}
\label{fourier-transform}
\tilde\psi(\bx,\bp,t) = {\int\!\!\int\!\!\int} \hat\psi(\bkappa,\bp,t) e^{i \bkappa\cdot\bx} \, d\bkappa
\end{equation}
It can be seen that if we consider only terms of order $\epsilon$ and lower, the different terms of the integrand do not interact.  As long as one assumes that $\bkappa$ evolves according to equation~\eqref{bkappa} (see below), without loss of generality, we can assume that the perturbation is of the form
\begin{equation}
\psi \to \psi + \epsilon\hat\psi e^{i \bkappa\cdot\bx}
\end{equation}
for some wave number $\bkappa$.  While this solution involves complex numbers, by considering linear combinations, we see that $e^{i \bkappa\cdot\bx}$ is simply `code' for $\sin(\bkappa\cdot\bx)$, and similarly $i e^{i \bkappa\cdot\bx}$ is the same solution that is $90^\circ$ out of phase, that is, $\cos(\bkappa\cdot\bx)$.  Note that this kind of perturbed quantity does not allow the boundary condition~\eqref{stokes-boundary} for $\bu$, but this is not a real problem, since we know that ultimately these terms come from equation~\eqref{fourier-transform}, and that $\tilde\psi$ does satisfy this boundary condition.  (From a more mathematical point of view, we are calculating the continuous spectrum of our linear operators rather than the point spectrum.)

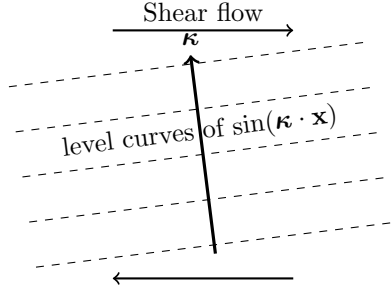
\begin{figure}
\begin{center}
\begin{tikzpicture}\begin{scope}[scale=0.3]
\draw[thick,->] (-4,5.5) -- node [above] {Shear flow} (4,5.5);
\draw[dashed] (-8.6,3) -- (7.4,5);
\draw[dashed] (-8.3,1) -- (7.7,3);
\draw[dashed] (-8,-1) -- node [above,sloped] {level curves of $\sin(\bkappa\cdot\bx)$} (8,1);
\draw[dashed] (-7.7,-3) -- (8.3,-1);
\draw[dashed] (-7.4,-5) -- (8.6,-3);
\draw [->,very thick](0.55,-4.4) -- (-0.55,4.4) node [above]{$\bkappa$} ;
\draw[thick,<-] (-4,-5.5) -- (4,-5.5) ;
\end{scope}\end{tikzpicture}
\end{center}
\caption{A diagram illustrating how $\bkappa$ is effected by a shear flow.  The level curves of $\sin(\bkappa\cdot\bx)$ will converge to the horizontal, and the wave number $\bkappa$ will converge to the vertical.}
\label{U kappa}
\end{figure}

In order to satisfy Jeffery's equation, we require that $\bkappa$ satisfy
\begin{equation}
\label{bkappa-evolves-so-that}
\left(\frac\partial{\partial t} + \bu\cdot\bnabla\right)(\hat\psi e^{i \bkappa\cdot\bx}) = \frac{\partial \hat \psi}{\partial t} \, e^{i\bkappa\cdot\bx}+O(\epsilon)
\end{equation}
To make this happen, we evolve $\bkappa$ according to the equation
\begin{gather}
\label{bkappa}
\frac\partial{\partial t}\bkappa = - \mathsf U^T \cdot \bkappa \\
\label{bkappa0}
\bkappa = \bkappa_0 \text{ at } t=0
\end{gather}
Equation~\eqref{bkappa-evolves-so-that} then follows because
\begin{equation}
\frac\partial{\partial t} e^{i\bkappa\cdot\bx}
=
-i[(\mathsf U^T\cdot\bkappa)\cdot\bx] e^{i\bkappa\cdot\bx}
=
-(\mathsf U\cdot\bx)\cdot(i\bkappa) e^{i\bkappa\cdot\bx}
=
-\bu\cdot\bnabla [e^{i\bkappa\cdot\bx}]
+O(\epsilon)
\end{equation}
Note that if $\mathsf U$ does not depend upon $t$, then equations~\eqref{bkappa} and~\eqref{bkappa0} are solved by
\begin{equation}
\label{bkappa-exp}
\bkappa = e^{-t \mathsf U^T}\bkappa_0
\end{equation}
We illustrate how a wave number $\bkappa$ is effected by a shear flow in Figure~\ref{U kappa}.

Now the perturbation in $\psi = \psi_{\mathsf B}$ is more easily described by a perturbation in $\mathsf B$:
\begin{equation}
\label{perturb B}
\mathsf B \to \mathsf B + \epsilon \hat{\mathsf B} e^{i\bkappa\cdot\bx}
\end{equation}
However, if we are going to measure the size of the growth of the perturbation, then $\epsilon\hat{\mathsf B}$ is not the correct thing to measure.  This is because $\epsilon \hat{\mathsf B}$ is, in some sense, an absolute error for $\mathsf B$.  Thus if $\epsilon\hat{\mathsf B}$ becomes large, this might simply reflect that $\mathsf B$ is becoming large.

For this reason, we instead measure a relative error, $\epsilon \tilde{\mathsf B}$, where
\begin{equation}
\label{tilde B}
\tilde{\mathsf B} = (\mathsf C^T)^{-1} \cdot \hat{\mathsf B} \cdot \mathsf C^{-1}
\end{equation}
This also simplifies calculations, because $\text{Tr}(\tilde{\mathsf B}) = 0$, a consequence that the perturbation needs to preserve $\text{det}(\mathsf B)=1$.

The goal, then, is to create an equation
\begin{equation}
\label{matrix-ode}
\frac{\partial \tilde{\mathsf B}}{\partial t} = M(\bkappa,t)[\tilde{\mathsf B}]
\end{equation}
where $M(\bkappa,t)$ is linear operator acting upon symmetric, trace zero matrices, depending (amongst other things) upon $\bkappa$.

The solution to equation~\eqref{matrix-ode} is described by the linear operator $L(\bkappa_0,t)$, where
\begin{equation}
\label{matrix-ode-soln}
\tilde{\mathsf B}(t) = L(\bkappa_0,t) [\tilde{\mathsf B}(0)]
\end{equation}
Here $\bkappa_0$ is as defined in equation~\eqref{bkappa0}, and we assume that $\bkappa$ in equation~\eqref{matrix-ode} satisfies equation~\eqref{bkappa}.

The largeness of the operator $L(\bkappa_0,t)$ can be found in two possible ways, by computing its spectral norm (that is the largest singular value), or by computing its spectral radius (that is the largest absolute value of its eigenvalues).  The second approach always gives a smaller or equal answer than the first approach, and so we can regard the second approach as the more conservative.  We use both approaches in our calculations, and we will see that in our specific situations that the answers are not significantly different.

\section{The Linearized Equations}

We make the additional replacements
\begin{gather}
\bu \to \mathsf U\cdot\bx+\epsilon \hat \bu e^{i\bkappa\cdot\bx} \\
\label{perturb Gamma}
\mathsf \Gamma \to \mathsf \Gamma + \epsilon\hat{\mathsf \Gamma} e^{i\bkappa\cdot\bx} \\
\label{perturb Omega}
\mathsf \Omega \to \mathsf \Omega+\epsilon \hat{\mathsf \Omega} e^{i\bkappa\cdot\bx} \\
\mathsf \sigma \to \mathsf \sigma+\epsilon \hat{\mathsf \sigma} e^{i\bkappa\cdot\bx} \\
q \to q+\epsilon \hat q e^{i\bkappa\cdot\bx} \\
\mathbb A \to \mathbb A+\epsilon \hat{\mathbb A} e^{i\bkappa\cdot\bx}
\end{gather}
The linearized equations can be shown to be
\begin{gather}
\label{fourier-psi}
\frac{\partial \tilde{\mathsf B}}{\partial t} = - \tfrac12\mathsf C\cdot(\hat{\mathsf \Omega}+\lambda\hat{\mathsf \Gamma})\cdot \mathsf C^{-1} - \tfrac12(\mathsf C^T)^{-1} \cdot(- \hat{\mathsf \Omega}+\lambda\hat{\mathsf \Gamma})\cdot \mathsf C^T \\
\label{fourier-A}
\hat{\mathbb A} = \frac{\partial \mathbb A}{\partial \mathsf B} : (\mathsf C^T\cdot\tilde{\mathsf B}\cdot \mathsf C) \\
\label{fourier-gamma}
\hat{\mathsf \Gamma} = i(\bkappa \hat\bu + \hat\bu \bkappa) \\
\label{fourier-omega}
\hat{\mathsf \Omega} = i(\bkappa \hat\bu - \hat\bu \bkappa) \\
\label{fourier-stress-def}
\hat{\mathsf \sigma} = \beta \mathbb A:\hat{\mathsf \Gamma} + \beta \hat{\mathbb A}:\mathsf \Gamma + \hat{\mathsf \Gamma} -\hat q \mathsf I \\
\label{fourier-stress-equ}
\bkappa \cdot \hat{\mathsf \sigma} = 0 \\
\label{fourier-div-free}
\bkappa \cdot \hat\bu = 0 \\
\label{dA/dB}
\frac{\partial \mathbb A}{\partial \mathsf B}
= - \tfrac{15}8 \int_0^\infty \frac{s\,\mathcal S[(\mathsf B+s \mathsf I)^{-1}\otimes(\mathsf B+s \mathsf I)^{-1}\otimes(\mathsf B+s \mathsf I)^{-1}] \, ds}{\sqrt{\text{det}(\mathsf B+s \mathsf I)}}
\end{gather}
noting here that $\frac{\partial \mathbb A}{\partial \mathsf B}$ is a rank 6 tensor.

Let us provide some details of the derivation of equation~\eqref{fourier-psi}.  Substituting equations~\eqref{perturb B}, \eqref{perturb Gamma} and~\eqref{perturb Omega} into equation~\eqref{B-full}, and retaining only the terms of order $\epsilon$, we obtain
\begin{equation}
\begin{split}
\frac\partial{\partial t} (\hat{\mathsf B}\,  e^{i\bkappa\cdot\bx})
+
\bu\cdot\bnabla(\hat{\mathsf B}\,  e^{i\bkappa\cdot\bx})
=
& - \tfrac12\hat{\mathsf B} \cdot(\mathsf \Omega+\lambda \mathsf \Gamma) \,  e^{i\bkappa\cdot\bx}
- \tfrac12(-\mathsf \Omega+\lambda \mathsf \Gamma) \cdot \hat{\mathsf B} \,  e^{i\bkappa\cdot\bx} \\
& - \tfrac12 \mathsf B \cdot(\hat{\mathsf \Omega}+\lambda \hat{\mathsf \Gamma}) \, e^{i\bkappa\cdot\bx}
- \tfrac12 (-\hat{\mathsf \Omega}+\lambda \hat{\mathsf \Gamma})\cdot \mathsf B \, e^{i\bkappa\cdot\bx}
\end{split}
\end{equation}
Applying equations~\eqref{bkappa-evolves-so-that}, \eqref{C-full} and~\eqref{tilde B}, dividing by $e^{i\bkappa\cdot\bx}$, and neglecting terms of order $\epsilon$ and higher, we obtain
\begin{equation}
\mathsf C^T \cdot \frac{\partial \tilde{\mathsf B}}{\partial t} \cdot \mathsf C = - \tfrac12 \mathsf B \cdot(\hat{\mathsf \Omega}+\lambda \hat{\mathsf \Gamma}) - \tfrac12 (-\hat{\mathsf \Omega}+\lambda \hat{\mathsf \Gamma}) \cdot \mathsf B
\end{equation}
and equation~\eqref{fourier-psi} is established.

Next, setting
\begin{equation}
\mathsf N = |\bkappa|^2 \mathsf I + 2 \beta \mathbb A:\bkappa\bkappa
\end{equation}
we have that equations~\eqref{fourier-stress-def}, \eqref{fourier-stress-equ} and~\eqref{fourier-div-free} can be replaced by
\begin{equation}
\label{fourier-bu}
\hat\bu = i \beta \left(\mathsf I - \frac{\mathsf N^{-1} \cdot \bkappa\bkappa}{\mathsf N^{-1}:\bkappa\bkappa} \right)\cdot \mathsf N^{-1} \cdot (\hat{\mathbb A}:\mathsf \Gamma)\cdot\bkappa
\end{equation}
To see this, note first that $\mathsf N$ is positive definite, and hence invertible.  Substitute equations~\eqref{fourier-gamma} and~\eqref{fourier-stress-def} into \eqref{fourier-stress-equ}, and apply equation~\eqref{fourier-div-free}, to obtain
\begin{equation}
\label{app2}
\hat\bu + i \mathsf N^{-1} \cdot \bkappa \hat q = i \beta \mathsf N^{-1} \cdot (\hat{\mathbb A}:\mathsf \Gamma)\cdot\bkappa
\end{equation}
Dot producting both sides with $\bkappa$, and again applying equation~\eqref{fourier-div-free}, yields
\begin{equation}
\hat q = \beta \frac{\bkappa\cdot \mathsf N^{-1} \cdot (\hat{\mathbb A}:\mathsf \Gamma)\cdot\bkappa}{\mathsf N^{-1}:\bkappa\bkappa}
\end{equation}
Substituting this into equation~\eqref{app2} gives equation~\eqref{fourier-bu}.

Finally, it is important to note that in equation~\eqref{matrix-ode}, that $M(\bkappa,t)$ only depends on the direction of $\bkappa$, not on its length.

\section{Linear perturbations of the fourth moment tensor}

Equation~\eqref{dA/dB} follows, since by differentiating equation~\eqref{A from B} we have for any symmetric matrix $\hat{\mathsf B}$
\begin{equation}
\begin{split}
\frac{\partial \mathbb A}{\partial \mathsf B} : \hat{\mathsf B}
= {} &
\frac{d}{dt} \left[\mathbb A(\mathsf B+t \hat{\mathsf B})\right] \Big|_{t=0} \\
= {} & - \tfrac32 \int_0^\infty \frac{s\,\mathcal S[((\mathsf B+s \mathsf I)^{-1}\cdot \hat{\mathsf B}\cdot(\mathsf B+s \mathsf I)^{-1})\otimes(\mathsf B+s \mathsf I)^{-1}] \, ds}{\sqrt{\text{det}(\mathsf B+s \mathsf I)}} \\
& - \tfrac38 \int_0^\infty \frac{s\, \text{Tr}[(\mathsf B+s \mathsf I)^{-1}\cdot \hat{\mathsf B}]\,\mathcal S[(\mathsf B+s \mathsf I)^{-1}\otimes(\mathsf B+s \mathsf I)^{-1}]\, ds}{\sqrt{\text{det}(\mathsf B+s \mathsf I)}}
\end{split}
\end{equation}
and noting that for any symmetric matrix $\mathsf K$
\begin{equation}
\mathcal S(\mathsf K\otimes \mathsf K\otimes \mathsf K):\hat{\mathsf B}
=
\tfrac45 \mathcal S((\mathsf K\cdot \hat{\mathsf B}\cdot \mathsf K)\otimes \mathsf K)
+
\tfrac15 \text{Tr}(\mathsf K\cdot \hat{\mathsf B})\,\mathcal S(\mathsf K\otimes \mathsf K)
\end{equation}
To compute $\frac{\partial \mathbb A}{\partial \mathsf B}$ in practice, one first diagonalizes $\mathsf B$ using an orthogonal similarity matrix, so that $\mathsf B=\text{diagonal}(b_1,b_2,b_3)$.  In this case, the following kinds of quantities appear
\begin{equation}
\mathcal E_{m_1,m_2,m_3}^n =
\int_0^\infty \frac{s^n \, ds}{(b_1+s)^{m_1+\frac12}(b_2+s)^{m_2+\frac12}(b_3+s)^{m_3+\frac12}}
\end{equation}
where $m_1$, $m_2$, $m_3$ and $n$ are non-negative integers satisfying $n \le m_1+m_2+m_3$.  For example, it is shown in \cite{montgomery-smith:09b} that the 4th moment tensor is given by the formulae (here $i\ne j\ne k$)
\begin{gather}
\mathbb A_{iiii} = \tfrac34\mathcal E^1_{2\delta_{i1},2\delta_{i2},2\delta_{i3}} \\
\mathbb A_{iijj} = \tfrac14\mathcal E^1_{\delta_{i1}+\delta_{j1},\delta_{i2}+\delta_{j2},\delta_{i3}+\delta_{j3}} \\
\mathbb A_{ijkk} = 0
\end{gather}
where $\delta_{ij}$ denotes the Kronecker delta symbol.  To compute $\mathcal E_{m_1,m_2,m_3}^n$, they are Carlson forms of elliptic integrals \cite[][]{carlson:95} if $n=0$ and $m_1+m_2+m_3=0$ or $1$.  Furthermore, if $b_1$, $b_2$ and $b_3$ are distinct, then the other quantities can be calculated using the following types of relations (the only mildly difficult equality is the first, which requires a single application of integration by parts):
\begin{gather}
(m_1+\tfrac12)\mathcal E^{n+1}_{m_1+1,m_2,m_3} + (m_2+\tfrac12)\mathcal E^{n+1}_{m_1,m_2+1,m_3} + (m_3+\tfrac12)\mathcal E^{n+1}_{m_1,m_2,m_3+1} = (n+1)\mathcal E^n_{m_1,m_2,m_3} \\
\mathcal E^n_{m_1,m_2+1,m_3+1} = (b_2-b_3)^{-1}(\mathcal E^n_{m_1,m_2,m_3+1}-\mathcal E^n_{m_1,m_2+1,m_3}) \\
\mathcal E^n_{m_1+1,m_2,m_3+1} = (b_3-b_1)^{-1}(\mathcal E^n_{m_1+1,m_2,m_3}-\mathcal E^n_{m_1,m_2,m_3+1}) \\
\mathcal E^n_{m_1+1,m_2+1,m_3} = (b_1-b_2)^{-1}(\mathcal E^n_{m_1,m_2+1,m_3}-\mathcal E^n_{m_1+1,m_2,m_3}) \\
\mathcal E^{n+1}_{m_1+1,m_2,m_3} = \mathcal E^n_{m_1,m_2,m_3} - b_1\mathcal E^n_{m_1+1,m_2,m_3} \\
\mathcal E^{n+1}_{m_1,m_2+1,m_3} = \mathcal E^n_{m_1,m_2,m_3} - b_2\mathcal E^n_{m_1,m_2+1,m_3} \\
\mathcal E^{n+1}_{m_1,m_2,m_3+1} = \mathcal E^n_{m_1,m_2,m_3} - b_3\mathcal E^n_{m_1,m_2,m_3+1}
\end{gather}
From a numerical perspective, good approximations to these elliptic integrals can also be found when the eigenvalues are not distinct by artificially adding a small term to some of the eigenvalues to make them distinct.  To avoid excessive floating point errors, the eigenvalues need to be quite far from each other, for example, using IEEE double precision arithmetic we found that the eigenvalues have to be at least $10^{-6}$ from each other to get reasonable results when $m_1+m_2+m_3 \le 3$.

If $i\ne j\ne k \ne i$, we have the formulae
\begin{gather}
\label{dA/dB iiiiii}
\left(\frac{\partial \mathbb A}{\partial \mathsf B}\right)_{iiiiii}
= -\tfrac{15}8 \mathcal E^1_{3\delta_{i1},3\delta_{i2},3\delta_{i3}}
\\
\label{dA/dB iiiijj}
\left(\frac{\partial \mathbb A}{\partial \mathsf B}\right)_{iiiijj}
=
-\tfrac38 \mathcal E^1_{2\delta_{i1}+\delta_{j1},2\delta_{i2}+\delta_{j1},2\delta_{i3}+\delta_{j1}} \\
\label{dA/dB iijjkk}
\left(\frac{\partial \mathbb A}{\partial \mathsf B}\right)_{iijjkk}
=
-\tfrac18 \mathcal E^1_{\delta_{i1}+\delta_{j1}+\delta_{k1},\delta_{i2}+\delta_{j2}+\delta_{k2},\delta_{i3}+\delta_{j3}+\delta_{k3}}
\end{gather}
and all terms which have any index appearing precisely an odd number of times are 0.

If the eigenvalues are all the same, that is, $\mathsf B=\mathsf I$, then equation~\eqref{dA/dB} simplifies to the formula
\begin{equation}
\label{dA/dB 0}
\frac{\partial \mathbb A}{\partial \mathsf B} = - \tfrac3{14} \mathcal S(\mathsf I\otimes \mathsf I\otimes \mathsf I)
\end{equation}

\section{Calculating the growth of perturbations}

The linear operator $M(\bkappa,t)$ in equation~\eqref{matrix-ode} can be computed by combining the
equations~\eqref{BC},
\eqref{C-full},
\eqref{C0},
\eqref{A from B},
\eqref{Gamma},
\eqref{Omega},
\eqref{fourier-psi},
\eqref{fourier-A},
\eqref{fourier-gamma},
\eqref{fourier-omega},
\eqref{dA/dB},
\eqref{fourier-bu},
\eqref{dA/dB iiiiii},
\eqref{dA/dB iiiijj},
and~\eqref{dA/dB iijjkk}.
If we only want to compute $M(\bkappa,0)$, then we can use equation~\eqref{dA/dB 0} in place of \eqref{dA/dB iiiiii}, \eqref{dA/dB iiiijj}, and~\eqref{dA/dB iijjkk}.  Solving equation~\eqref{matrix-ode} to give $L(\bkappa_0,t)$, as described in equation~\eqref{matrix-ode-soln} requires the additional equations~\eqref{bkappa} and~\eqref{bkappa0}.

The operators $M(\bkappa,t)$ and $L(\bkappa_0,t)$ act on the space of symmetric, trace zero, $3\times 3$ matrices.  This is a five dimensional space, and with respect to the Frobenius norm, has an orthogonal basis
\begin{equation}
\begin{split}
\mathsf T_1 &= \tfrac1{\sqrt2}\left[\begin{smallmatrix}
1 & 0 & 0 \\
0 & -1 & 0 \\
0 & 0 & 0
\end{smallmatrix}\right]
,\quad
\mathsf T_2 = \tfrac1{\sqrt2}\left[\begin{smallmatrix}
0 & 1 & 0 \\
1 & 0 & 0 \\
0 & 0 & 0
\end{smallmatrix}\right]
,\quad
\mathsf T_3 = \tfrac1{\sqrt6}\left[\begin{smallmatrix}
1 & 0 & 0 \\
0 & 1 & 0 \\
0 & 0 & -2
\end{smallmatrix}\right]
,\\
\mathsf T_4 &= \tfrac1{\sqrt2}\left[\begin{smallmatrix}
0 & 0 & 1 \\
0 & 0 & 0 \\
1 & 0 & 0
\end{smallmatrix}\right]
,\quad
\mathsf T_5 = \tfrac1{\sqrt2}\left[\begin{smallmatrix}
0 & 0 & 0 \\
0 & 0 & 1 \\
0 & 1 & 0
\end{smallmatrix}\right]
\end{split}
\end{equation}
Thus $M(\bkappa,t)$ and $L(\bkappa_0,t)$ with respect to this basis are given by the five by five matrices $\mathcal M(\bkappa,t)$ and $\mathcal L(\bkappa_0,t)$ where
\begin{gather}
\mathcal M(\bkappa,t)_{i,j} = \text{Tr}(\mathsf T_i \cdot (M(\bkappa,t)[\mathsf T_j]))
\\
\mathcal L(\bkappa_0,t)_{i,j} = \text{Tr}(\mathsf T_i \cdot (L(\bkappa_0,t)[\mathsf T_j]))
\end{gather}
Thus we obtain the ordinary differential equation
\begin{gather}
\frac{\partial}{\partial t} \mathcal L(\bkappa_0,t) = \mathcal M(\bkappa,t) \cdot \mathcal L(\bkappa_0,t) \\
\mathcal L(\bkappa_0,0) = I
\end{gather}
The growth of $\mathcal L(\bkappa_0,t)$ is measured by either computing $n(t)$, the maximum over all $\bkappa_0 \in S$ of the spectral norm of $\mathcal L(\bkappa_0,t)$, or if we want to be more conservative, $r(t)$, the maximum over all $\bkappa_0 \in S$ of the spectral radius of $\mathcal L(\bkappa_0,t)$.

The calculations were performed with a C++ program, making use of the \emph{Newmat} software package \cite[][]{newmat} for the matrix calculations, and the \emph{GSL} software package \cite[][]{gsl} for the elliptic integrals.  Also, because the differential equation turned out to be rather stiff, we used the implicit Runge-Kutta method of order 5 (Radau IIA) by \cite{hairer:96b,hairer:96a} for solving the ordinary differential equation.  The C++ programs may be found at \url{http://www.math.missouri.edu/~stephen/software/jeff-stokes}.

\section{Demonstration of `buckling'}
\label{section buckling}

Here the idea is to see how fast $\mathcal L(\bkappa_0,t)$ grows for small $t\ge0$.  This can be performed by computing the eigenvalues of $\mathcal M(\bkappa,0)$ for various values of $\bkappa$, and to see if any of them have positive real part.  Having computed the eigenvalues of $M(\bkappa,0)$, we feed them into equation~\eqref{fourier-bu} to see if they produce a significant perturbation to $\bu$, and to see in what direction these perturbations take place.

\begin{figure}
\begin{center}
\psfrag{phi}{$\phi$ in degrees}
\psfrag{theta}{$\theta$ in degrees}
\psfrag{m(theta,phi)}{$m(\theta,\phi)$}
\includegraphics[trim=4cm 3cm 4cm 2.5cm,clip]{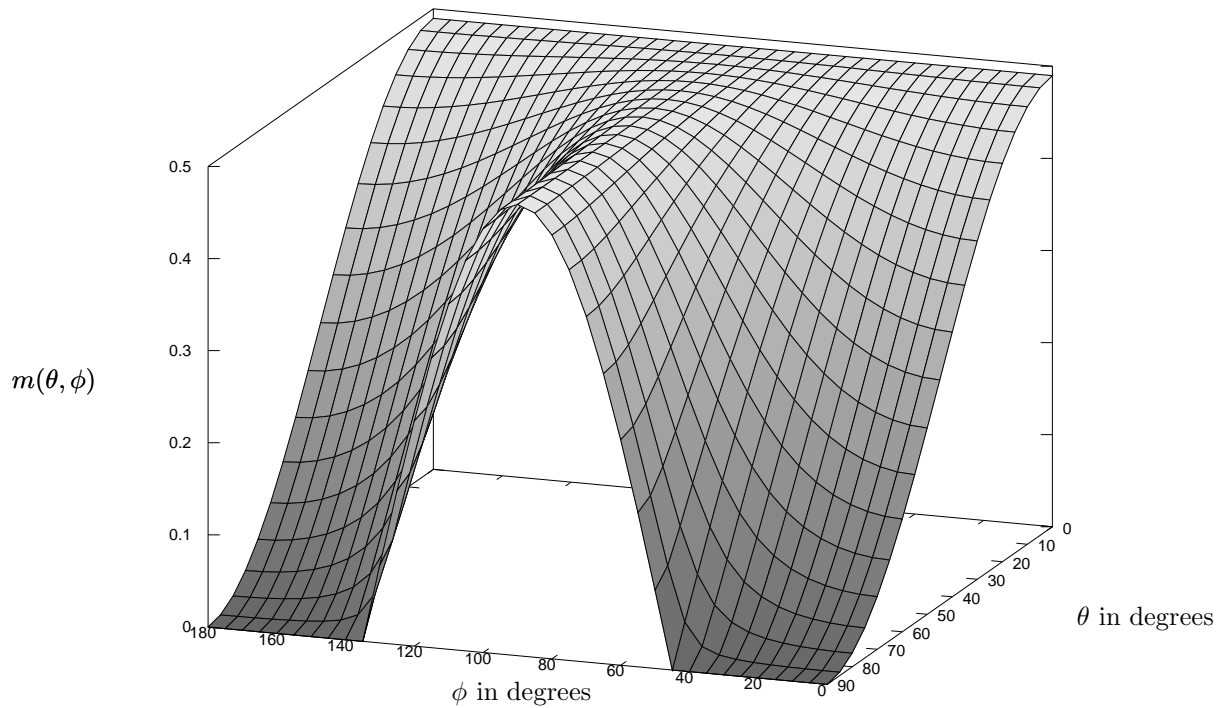}
\caption{Graph of largest real part of eigenvalue of $\mathcal M(\bkappa,0)$ under elongation flow.}
\label{fig1}
\end{center}
\end{figure}

In Figure~\ref{fig1} we display $m(\theta,\phi)$, the largest real part of the eigenvalues of $\mathcal M(\bkappa,0)$, with elongation flow $\bnabla\bu=\left[\begin{smallmatrix}1&0&0\\0&-1&0\\0&0&0\end{smallmatrix}\right]$ and $\lambda=1$ and $\beta=10$.  This was computed using the software product \emph{Mathematica} by Wolfram Research, Inc., and independently verified by the software described in the next section.  We only need to consider $\bkappa$ on the unit sphere, that is, when $\bkappa = (\cos(\phi)\sin(\theta),\sin(\phi)\sin(\theta),\cos(\theta))$, with $0<\phi<\pi$ and $0<\theta<\pi/2$.

We see that this becomes particularly large, with value $24/49$, when $\phi = 90^\circ$.  For example, if $(\theta,\phi)=(90^\circ,90^\circ)$, then equation~\eqref{fourier-bu} gives $\hat\bu = (0,0, 24/49)$.  This is exactly the kind of buckling illustrated in Figure~\ref{buckling}.  When $\theta=0$, we find that $\hat\bu = \pm(0,24/49, 0)$, that is, the fibers slide past each other in the $xy$-plane in a direction parallel to the $y$-axis.

Notice that the perturbations use all three dimensions.  Simulations of the coupled system have usually been performed in only two dimensions \cite[][]{chinesta:02,chinesta:03}, or at least with some assumption of symmetry that effectively limits the spacial degrees of freedom to two dimensions \cite[][]{verweyst:02}.  So we also performed the computation assuming that the fluid has no freedom to move in the $z$-direction.  This is achieved by limiting $\hat{\mathsf C}$ so that its non-zero components are only in the top left two by two submatrix (equivalently considering only the basis elements $\mathsf T_1$ and $\mathsf T_2$, or only considering the top left two by two submatrix of $\mathcal M(\bkappa,0)$), and setting $\theta=90^\circ$.  Computations reveal that $M(\bkappa,0) = 0$.  This suggests that to obtain a true sense of how instabilities effect the flow, simulations should allow freedom in all three dimensions, with no symmetry assumptions.

\section{Analysis of growth of perturbations}
\label{section analysis of growth}

We denote by $n(t)$ the maximum of the spectral norm of $\mathcal L(\bkappa_0,t)$ over all $\bkappa_0$ in the two dimensional sphere, and by $r(t)$, the maximum over all $\bkappa_0$ of the spectral radius of $\mathcal L(\bkappa_0,t)$.  Again we are looking at the case with shear flow $\bnabla\bu=\left[\begin{smallmatrix}0&1&0\\0&0&0\\0&0&0\end{smallmatrix}\right]$ and $\lambda=1$ and $\beta=10$, $100$ and $1000$.  The results are shown in Table~\ref{growth-beta}.  Since we are only interested in orders of magnitude, the spectral norms are only calculated to about two digits of accuracy.  It can be seen that it takes quite a while ($t \approx 6$) before the growths even begin to become large enough that we might expect significant difference from the unperturbed solution.  It may also be observed that the values of the spectral radius are of the same magnitude as the values of the spectral norm.

\begin{table}
\begin{center}
\begin{tabular}{@{}ccccccc@{}}
&
\multicolumn{2}{c}{$\beta=10$} &
\multicolumn{2}{c}{$\beta=100$} &
\multicolumn{2}{c}{$\beta=1000$} \\
$t$
& $n(t)$ & $r(t)$
& $n(t)$ & $r(t)$
& $n(t)$ & $r(t)$ \\
1 & 1.2 & 1.2 & 1.3 & 1.3 & 1.3 & 1.3 \\
2 & 1.3 & 1.1 & 1.6 & 1.2 & 1.7 & 1.2 \\
3 & 2.1 & 1.3 & 2.7 & 1.6 & 2.7 & 1.6 \\
4 & 4 & 2.1 & 6 & 2.8 & 6.4 & 2.8 \\
5 & 6.5 & 3.7 & 12 & 6.4 & 13 & 6.9 \\
6 & 9 & 5.9 & 19 & 13 & 22 & 15 \\
7 & 11 & 8.1 & 30 & 24 & 38 & 30 \\
8 & 12 & 9.8 & 48 & 42 & 68 & 59 \\
9 & 12 & 11 & 83 & 70 & 140 & 110 \\
10 & 11 & 11 & 150 & 120 & 320 & 240 \\
\end{tabular}
\end{center}

\

\caption{Maximal growth of spectral norm and spectral radius of linearized perturbations starting with isotropic data.}
\label{growth-beta}
\end{table}

One can actually follow where the large growths take place, by analyzing the singular decomposition of $\mathcal L(\bkappa_0,t)$.  For example, for the case $t=10$ and $\beta=1000$, most of the growth takes place around $\bkappa$ in the direction $(\theta,\phi) = (90^\circ,-78.4^\circ)$ (having coming from $\bkappa_0$ directed $(\theta,\phi) = (90^\circ,79^\circ)$), and $\hat\bu$ is directed parallel to the $z$-axis.  This looks like an extreme form of the `buckling' effect described above.

\section{Different Initial Conditions}
\label{section different initial}

The actual physical experiments can be performed by pouring the suspension upon a plate, and then running another plate along the top.  It is to be expected (and has been reported) that the initial fiber distribution is not isotropic, but rather the effect of the suspension being squeezed by the pouring action is that most of the fibers are aligned in the $xz$-plane.  Squeezing the fluid $\rho>1$ times can be simulated by a fluid gradient $\bnabla \bu = \left[\begin{smallmatrix}1/2 & 0 & 0 \\ 0 & -1 & 0 \\ 0 & 0 & 1/2\end{smallmatrix}\right]$ running for a time $\log(\rho)$.  Thus, this squeezing can be modeled by replacing equation~\eqref{C0} by
\begin{equation}
\mathsf C = \mathsf C_\rho \text{ at } t=0
\end{equation}
where
\begin{equation}
\mathsf C_\rho = \left[\begin{matrix}\rho^{-\lambda/2} & 0 & \\ 0 & \rho^\lambda & 0 \\ 0 & 0 & \rho^{-\lambda/2}\end{matrix}\right]
\end{equation}
and in the case that $\mathsf U$ is independent of $t$, equation~\eqref{C-exp} is replaced by
\begin{equation}
\mathsf C = \mathsf C_\rho \cdot \exp\left(-\tfrac12 t (\mathsf \Omega + \lambda\mathsf \Gamma)\right)
\end{equation}

In Table~\ref{growth-s} we show growth of perturbations for the shear flow $\bnabla\bu=\left[\begin{smallmatrix}0&1&0\\0&0&0\\0&0&0\end{smallmatrix}\right]$ and $\lambda=1$ and $\beta = 100$ and $\rho=1$, $2$, $3$ and $4$.

\begin{table}
\begin{center}
\begin{tabular}{@{}lcccccccc@{}}
&
\multicolumn{2}{c}{$\rho=1$} &
\multicolumn{2}{c}{$\rho=2$} &
\multicolumn{2}{c}{$\rho=3$} &
\multicolumn{2}{c}{$\rho=4$} \\
$t$
& $n(t)$ & $r(t)$
& $n(t)$ & $r(t)$
& $n(t)$ & $r(t)$
& $n(t)$ & $r(t)$ \\
0.5 & 1.2 & 1.2 & 1.7 & 1.4 & 3 & 1.9 & 4.6 & 2.6 \\
1 & 1.3 & 1.3 & 2.6 & 1.8 & 6.9 & 3.6 & 14 & 6.7 \\
1.5 & 1.4 & 1.3 & 3.9 & 2.3 & 14 & 6.2 & 39 & 16 \\
2 & 1.6 & 1.2 & 5.4 & 2.7 & 27 & 10 & 100 & 38 \\
2.5 & 1.9 & 1.4 & 7.2 & 3 & 49 & 16 & 240 & 84 \\
3 & 2.7 & 1.6 & 9.3 & 3.2 & 84 & 25 & 570 & 180 \\
3.5 & 4 & 1.8 & 12 & 3.2 & 140 & 36 & 1300 & 360 \\
4 & 6 & 2.8 & 14 & 3.4 & 210 & 51 & 2700 & 680 \\
4.5 & 8.5 & 4.2 & 17 & 4.7 & 310 & 67 & 5400 & 1300 \\
5 & 12 & 6.4 & 19 & 6.7 & 430 & 81 & 10000 & 2200 \\
\end{tabular}
\end{center}

\

\caption{Growth of spectral properties with initial data created by squeezing.}
\label{growth-s}
\end{table}

Another way the experiment can be performed (see \cite{wang:08}) is to first apply a reverse shear to the suspension until it achieves its steady state, and then to use this as the initial condition.  Running a shear in the opposite direction using the Folgar-Tucker equation \cite[][]{folgar:84} with $C_I=0.01$ gives a second order tensor matrix
\begin{equation}
\mathsf A = \left[\begin{matrix}
0.727123 & -0.074765 & 0 \\
-0.074765 & 0.0923099 & 0 \\
0 & 0 & 0.180567 \end{matrix}\right]
\end{equation}
(this was computed using spherical harmonics \cite[][]{spherical,montgomery-smith:09a}), which by inverting the elliptic integrals \cite[see][]{montgomery-smith:09b}, can be shown to correspond to
\begin{equation}
\mathsf C = \left[\begin{matrix}
0.372914 & 0.216212 & 0 \\
0.216212 & 2.20872 & 0 \\
0 & 0 & 1.28714
\end{matrix}\right]
\end{equation}
(Note that $\mathsf C$ is uniquely determined from $\mathsf B$ up to pre-multiplication by an orthogonal matrix, and pre-multiplying $\mathsf C$ by an orthogonal matrix has the effect of changing $\tilde{\mathsf B}$ by an orthogonal change of basis, which does not change its spectral norm.)  Using this value of $\mathsf C$ as the initial data, and applying the above shear with $\lambda=1$ and $\beta=100$ gave the results shown in Table~\ref{growth-preshear}.

\begin{table}
\begin{center}
\begin{tabular}{@{}lcc@{}}
$t$ & $n(t)$
& $r(t)$ \\
0.5 & 4.7 & 2.9 \\
1 & 14 & 6.8 \\
1.5 & 37 & 13 \\
2 & 81 & 22 \\
2.5 & 160 & 29 \\
3 & 280 & 43 \\
3.5 & 470 & 130 \\
4 & 720 & 290 \\
4.5 & 1100 & 540 \\
5 & 1500 & 900 \\
\end{tabular}
\end{center}

\

\caption{Growth of spectral properties with initial data created by a reverse shear.}
\label{growth-preshear}
\end{table}

It can be seen that both sets of initial conditions illustrated here give rise to growths of perturbations that greatly exceed the growth with isotropic initial conditions.

\section{Conclusion}
\label{section conclusion}

The results from Tables~\ref{growth-beta}, \ref{growth-s} and~\ref{growth-preshear} show that the size of $\tilde{\mathsf B}$ increases by a fairly substantial amount.  If the original perturbations are reasonably large, one can see that even after a short amount of time, that large deviations from the unperturbed solution are very likely.  We propose that this is what leads to the greatly reduced rate of alignment of the fibers.

We suspect that the perturbations take place at small length scales, in both numerical and physical experiments.  The reason we think this is as follows.  The rate of growth of the perturbations does not depend upon the size of the wave number.  One of our assumptions is that the medium is of infinite extent.  If we have the more realistic situation where there are boundaries to the medium, it is reasonable to think that if the wave number is smaller (that is, the length scale of the oscillations of the perturbations is larger), that the boundary has a greater effect, and that most likely this effect slows down the growth.  Thus we think that the perturbations take place with large wave numbers.  The largest realistic wave number for physical experiments is probably the reciprocal of the length of the fibers, and in numerical experiments is probably the reciprocal of the size of the spacial grid or elements used.  For this reason, we suspect that the growth of perturbations may be quite difficult to measure in physical experiments, and in numerical experiments might be mistaken for problems with the numerical method.  Indeed, when one is operating at these kinds of small length scales, one begins to doubt even the assumption that discrete fibers can be well represented by the continuum.  This perhaps undermines the original assumption of this paper, that the complex hydrodynamic interactions between the fibers can be somehow modeled by continuum equations.  Nevertheless, we still feel that there is a good chance that the coupled continuum equations might accurately describe the evolution of flows of fibers in fluids.

For future work, we would like to see the result of a full numerical simulation of the coupled Jeffery-Stokes equation, that is,
equations~\eqref{stokes},
\eqref{incompressible},
\eqref{stress},
\eqref{B-full}
and~\eqref{A from B},
replacing equation~\eqref{B0} with $\mathsf B$ initially set to a small perturbation of $\mathsf I$.  We note that some numerical simulations of the coupled equations have already been performed \cite{chinesta:02,chinesta:03,verweyst:02}, but we believe that more detailed computations, that allow complete freedom in all three spacial dimensions, are necessary to observe the effects predicted in this paper.

\section{Acknowledgments}

The author gratefully acknowledges support from N.S.F.\ grant C.M.M.I.\ 0727399.  The author also is grateful for comments from the reviewers of the paper, which were extremely helpful.

\bibliographystyle{jfm}
\bibliography{info_08}

\section{Corrigendum}

While Shibi Vasudeva and the author were performing numerical experiments to verify the predictions made in in this paper, it was found that it contained an error.  Equation~\eqref{fourier-omega} ((6.10) in the published version) contains a sign error, and should read
\begin{equation*}
\tilde{\mathsf \Omega} = i(\hat\bu\bkappa-\bkappa\hat\bu)
\end{equation*}
Also, the author decided not to report the spectral radius of $\mathcal L(\bkappa_0,t)$ in Sections~\ref{section analysis of growth} and~\ref{section different initial}.  This is because in equation~\eqref{tilde B}/(5.10), $\mathsf C$ could conceivably be replaced by any matrix that satisfies equation~\eqref{psi-B}/(3.2), and hence is only uniquely defined up to left multiplication by an orthogonal matrix.  Hence $\mathcal L(\bkappa_0,t)$ is unique only up to left multiplication by the five by five orthogonal matrix that represents conjugation by an orthogonal matrix on the space of three by three trace zero matrices.  This does not affect the spectral norm.

The results in Section~\ref{section buckling} remain unchanged.  The results in Sections~\ref{section analysis of growth} and~\ref{section different initial} are different.  Tables~\ref{growth-beta}, \ref{growth-s} and~\ref{growth-preshear} should be replaced by Tables~\ref{growth-beta-new}, \ref{growth-s-new} and~\ref{growth-preshear-new}, respectively.  The statement in Section~\ref{section analysis of growth} regarding where the large growths take place should be disregarded.  The conclusions in Section~\ref{section conclusion} are probably still valid, but the arguments in their favor are less compelling.

\begin{table}
\begin{center}
\begin{tabular}{@{}cccc@{}}
&
$\beta=10$ &
$\beta=100$ &
$\beta=1000$ \\
$t$
& $n(t)$
& $n(t)$
& $n(t)$ \\
1 & 1.2 & 1.3 & 1.2 \\
2 & 1.7 & 2.1 & 2.1 \\
3 & 3.3 & 4.8 & 5 \\
4 & 6.7 & 11 & 12 \\
5 & 12 & 22 & 24 \\
6 & 21 & 48 & 54 \\
7 & 25 & 43 & 49 \\
8 & 42 & 91 & 68 \\
9 & 51 & 170 & 200 \\
10 & 28 & 98 & 120 \\
\end{tabular}
\end{center}

\

\caption{Maximal growth of spectral norm and spectral radius of linearized perturbations starting with isotropic data.}
\label{growth-beta-new}
\end{table}

\begin{table}
\begin{center}
\begin{tabular}{@{}lcccc@{}}
&
$\rho=1$ &
$\rho=2$ &
$\rho=3$ &
$\rho=4$ \\
$t$
& $n(t)$
& $n(t)$
& $n(t)$
& $n(t)$ \\
0.5 & 1.2 & 1.6 & 2.3 & 3 \\
1 & 1.3 & 1.9 & 3.2 & 4.4 \\
1.5 & 1.5 & 2.3 & 3.9 & 5.6 \\
2 & 2.1 & 2.8 & 4.7 & 6.8 \\
2.5 & 3.1 & 3.6 & 5.5 & 8 \\
3 & 4.8 & 4.7 & 6.6 & 9.4 \\
3.5 & 7.2 & 6.2 & 8.1 & 9.8 \\
4 & 11 & 8.3 & 9.3 & 11 \\
4.5 & 16 & 9.8 & 12 & 14 \\
5 & 22 & 12 & 14 & 17 \\
\end{tabular}
\end{center}

\

\caption{Growth of spectral properties with initial data created by squeezing.}
\label{growth-s-new}
\end{table}

\begin{table}
\begin{center}
\begin{tabular}{@{}lc@{}}
$t$ & $n(t)$ \\
0.5 & 3.2 \\
1 & 4.8 \\
1.5 & 6.1 \\
2 & 7.8 \\
2.5 & 10 \\
3 & 13 \\
3.5 & 15 \\
4 & 18 \\
4.5 & 20 \\
5 & 22 \\
\end{tabular}
\end{center}

\

\caption{Growth of spectral properties with initial data created by a reverse shear.}
\label{growth-preshear-new}
\end{table}

\end{document}